\documentclass[final,3p,twocolumn,times]{elsarticle}

\makeatletter
\def\ps@pprintTitle{%
     \let\@oddhead\@empty
     \let\@evenhead\@empty
     \def\@oddfoot
     {\hbox to \textwidth%
     {\@myfooterfont%
     \begin{tabular}{l}
     \\\url{https://doi.org/10.1016/j.is.2025.102574}\\
     Accepted in Information Systems. [17 May 2025]\\ This manuscript version is made available under the CC-BY 4.0 license\\
     \url{https://creativecommons.org/licenses/by/4.0/}.\end{tabular}%
     }
     }
     \let\@evenfoot\@oddfoot}
\makeatother

\usepackage{amssymb}
\usepackage{amsmath}

\usepackage{multicol}
\usepackage{enumitem}
\usepackage{mathtools}
\usepackage{booktabs}
\usepackage{xcolor}
\usepackage[colorlinks=true]{hyperref}

\journal{Information Systems}

\begin{document}

\begin{frontmatter}



\title{PARK: Personalized Academic Retrieval with Knowledge-graphs}

\author[unimib,cnr]{Pranav Kasela\corref{cor1}} 
\ead{pranav.kasela@unimib.it}

\author[unimib]{Gabriella Pasi}
\ead{gabriella.pasi@unimib.it}

\author[cnr]{Raffaele Perego}
\ead{raffaele.perego@isti.cnr.it}

\cortext[cor1]{Corresponding author}

\affiliation[unimib]{organization={University of Milan-Bicocca},
            city={Milan},
            country={Italy}}
\affiliation[cnr]{organization={ISTI-CNR},
            city={Pisa},
            country={Italy}}

\begin{abstract}

Academic Search is a search task aimed to manage and retrieve scientific documents like journal articles and conference papers. Personalization in this context meets individual researchers’ needs by leveraging, through user profiles, the user related information (e.g. documents authored by a researcher), to improve search effectiveness and to reduce the information overload. While citation graphs are a valuable means to support the outcome of recommender systems, their use in personalized academic search (with, e.g. nodes as papers and edges as citations) is still under-explored.

Existing personalized models for academic search often struggle to fully capture users’ academic interests. To address this, we propose a two-step approach: first, training a neural language model for retrieval, then converting the academic graph into a knowledge graph and embedding it into a shared semantic space with the language model using translational embedding techniques. This allows user models to capture both explicit relationships and hidden structures in citation graphs and paper content.
We evaluate our approach in four academic search domains, outperforming traditional graph-based and personalized models in three out of four, with up to a 10\% improvement in MAP@100 over the second-best model. This highlights the potential of knowledge graph-based user models to enhance retrieval effectiveness.

\end{abstract}



\begin{keyword}


Personalized Information Retrieval \sep Knowledge Graphs \sep Neural Models
\end{keyword}

\end{frontmatter}

\section{Introduction}
\label{sec:introduction}
\textit{Academic search} is a task that focuses on retrieving scientific documents (e.g., research papers, articles, PhD manuscripts) from large repositories, based on queries, generally, formulated by researchers, students, educators, and professionals seeking specific information. Enhancing this process through search personalization offers the potential to tailor results to individual expertise and preferences, going beyond the \textit{one size fits all} approach of traditional Information Retrieval (IR) systems. Personalized Information Retrieval (PIR) systems make use of user models, which encode user-specific information and contextual cues to improve the assessment of relevance of search results.

The focus of this paper is on user modeling for personalized academic search, a task that remains under-explored in the literature. Specifically, our contributions are as follows:
    We propose a user model for academic search that leverages a knowledge graph constructed from citation data to capture user preferences and document relationships.
    Furthermore, we develop a complete retrieval pipeline, starting from an academic citation graph to create a personalized information retrieval system for academic search.

\textit{Citation graphs}, where nodes are academic papers and directed edges denote citations, can help enhance academic search by improving relevance assessment based on citation patterns and relationships between scientific documents.
While citation graphs have been successfully applied in academic recommender systems \cite{ali20graph}, their potential in PIR remains partially under-explored \cite{lu20knowledge}. 
To address this issue, in this paper, we propose a novel approach that uses knowledge graph embeddings to generate user models for personalized academic search. These embeddings map entities and relationships into a continuous vector space, capturing both explicit connections (e.g., citations) and latent structures (e.g., topical similarities and social connections) within citation graphs. 

Our approach integrates the principles of collaborative filtering by modeling user preferences based on relationships derived from the citation network.
By embedding papers and authors into a shared semantic space with a neural language model, our method enables user models to leverage both the structural and textual information of academic content. Specifically, the proposed approach aligns knowledge graph embeddings with neural language models through a two-stage training process. 
First, we train a bi-encoder neural language model on an academic search dataset. In the second step, we use a translational algorithm for knowledge graph embeddings to embed the knowledge graph, derived from the academic citation network, into the same semantic space as the neural language model. 
The two-stage training approach offers several advantages: it allows for seamless integration with any dense encoder with the knowledge graph without requiring retraining of the language model. Additionally, it allows the acceleration of the training process by pre-indexing document embeddings for efficient retrieval. 

The results of the performed experiments show that user models based on such knowledge graph significantly enhance retrieval effectiveness compared to other methods, including graph.based methods, achieving state-of-the-art performance in three out of four benchmarks in personalized academic search \cite{bassani22amulti}, with improvements of up to $10\%$ over the best-performing baseline.
By leveraging the semantic and structural information embedded within citation networks for user modeling, this work represents a significant advancement in personalized academic retrieval, offering a robust framework for future research and applications in this domain.

The work is organized as follows: In Section \ref{sec:related_work} we describe the training techniques employed for creating text embeddings and knowledge graph embeddings. In Section \ref{sec:methodology}, we propose PARK, an architecture that relies on two-stage training and embeds the knowledge graph in a semantic space shared with neural language models. The experimental settings are reported in Section \ref{sec:experimental_settings}, and the results are presented and discussed in Section \ref{sec:results}. In Section \ref{sec:ablation} we present an ablation study on the type of nodes utilized in the knowledge graph. Finally, Section \ref{sec:conclusion} concludes the work by providing new insights and new future research directions.

\section{Related Work and Background}
\label{sec:related_work}

Neural Information Retrieval (NIR) marks a major advancement in IR by using deep learning to enhance retrieval effectiveness \cite{bhaskar18anintroduction}. Unlike traditional IR models that depend on manual feature engineering and basic mathematical models, NIR systems automatically learn complex representations and relationships between queries and documents through deep learning.
In this work, we focus on the \textit{Siamese Networks} or Bi-encoders class of neural IR approaches  \cite{reimers19sentence}. 
The training of these models is done following the approach presented by Huang et al. \cite{huang13learning}, which minimizes the triplet margin loss \cite{balntas16learning} on triplet samples consisting of a query $q$, a relevant document $d^+$, and negative documents $D^-$ randomly sampled from the collection.
For efficiency purposes, the negative samples are taken from the samples used in the same batch, a technique known as random in-batch negatives \cite{karpukhin20dense,zhan20repbert}.

Personalized Information Retrieval (PIR) is a growing research field, driven by the need to enhance the relevance of search results for individual users \cite{tan06mining, sieg07websearch, carman10towardsquery, sontag12probabilistic}. 
Early efforts in personal search focused on leveraging metadata \cite{bendersky17learning}, contextual information like query time and location \cite{qin20matching, zamani17situational}, and developing training schemes to handle sparse user interactions \cite{bi21leveraging, wang16learning}.
Central to PIR is the definition of user models that capture various aspects of user behavior, preferences, and context to improve the relevance assessment process for specific user.
Initial approaches to building user representations included term-level models \cite{teeVan05personalizingPIR}, topic models \cite{sontag12probabilistic}, and latent representations learned through techniques like matrix factorization \cite{cai14personalized}.
The development of deep learning has significantly impacted also the definition of more accurate user models. Existing techniques rely on representation learning \cite{bengio13representation} to learn latent semantic representations of queries, documents, and user preferences for enhancing personalized search effectiveness \cite{ai19azero, zhong20personalized, yao20rlper}. 
More recent advancements include learning personalized word embeddings \cite{yao20personalwordembed}, and employing Recurrent Neural Networks (RNNs) \cite{ai19azero, ai17learninghierarchical}  and Transformers \cite{bi20transformer, zhong20personalized}  to model user representations.
A line of research has focused on leveraging hierarchical recurrent neural networks (HRNN) with query-aware attention to capture both short-term and long-term dependencies within user search sessions \cite{songwei18personalizing}. 
By integrating query-aware attention, these models can dynamically adjust the focus based on the specific query, leading to more accurate and more robust user profiles.
Additionally, graph neural networks \cite{zhan21modeling} and reinforcement learning techniques \cite{yao20rlper} have been explored to model complex user behaviors. 
Lu et al. \cite{lu20knowledge} propose a method that relies on a knowledge graph for personalized entity linking to improve search results based on user feedback during search sessions. In contrast to purely neural methods, such knowledge-enhanced personalization offers greater ability to capture structural relations across authors and topics.
These methods, particularly those with query-aware attention, have shown promising results in capturing the temporal dynamics of user interactions.
Despite advancements, these approaches often rely on the availability of session-based data and focus on personalization within search sessions, thus possibly failing to capture users' long-term interests. Furthermore, the reproducibility of results remains a challenge due to the use of proprietary datasets or to a lack of clear instructions for reproducing results \cite{bassani24denoising}. Bassani et al. specifically highlight that many personalized models lack public datasets or open-source code, hindering replicability and comparison.
Another research direction explores the use of attention mechanisms to enhance long-term user modeling during query time. Attention mechanisms enable the system to dynamically weigh and aggregate user-related information in relation to the current query, thus refining the personalization process \cite{bassani24denoising}. This method has proven effective in identifying the most relevant aspects of a user’s search history related to the given query.
Recently, the aspect of controllable personalization has gained attention, addressing concerns about the opacity of personalized systems and the lack of user control. Work in this area has explored providing users with interactive control over personalized search by making user representations, such as term or entity-based profiles, editable through visualization interfaces \cite{zemede17editableprofiles, ahn15openusermodels}. This allows users to inspect and modify how the system perceives their interests, aiming to improve transparency and user satisfaction. Recent cross-encoder based models like CtrlCE \cite{mysore25bridging} further extend this idea by integrating memory-augmented user profiles and a calibrated control mechanism that decides when personalization is beneficial. This line of work contributes to a growing interest in transparency, fairness, and user agency in IR \cite{chien23fairness}.

\subsection{Knowledge Graph Embeddings}

A knowledge graph (KG) is a data structure that encodes the knowledge of the real world in a graph format, whose nodes represent the entities and whose edges are the relationships that connect these entities. Depending on the application, KGs can be modeled as a directed edge-labelled graph, a heterogeneous graph, or a property graph \cite{hogan21knowledge}. 
Deep learning techniques can create dense representations of the graph in a continuous, low-dimensional vector space. 
One of the most commonly used families of knowledge graph embedding techniques is the translational model, which maps entities and relations in a knowledge graph to a continuous vector space, capturing semantic relationships and preserving their geometry. Among these translational algorithms, \textit{TransE} and \textit{TransH} are particularly popular due to their simplicity and effectiveness.

TransE, Translating Embeddings, was introduced by Bordes et al. \cite{bordes13transe}. It operates under the principle that relationships between entities can be represented as translations in the embedding space. Given a triplet $<h,r,t>$ in the knowledge graph, where h is the head entity, r is the relation, and t is the tail entity, TransE embeds these entities and relations into a continuous d-dimensional vector space.
The training process involves minimizing a margin-based ranking loss function, which encourages correct triplets to have a smaller distance than incorrect ones.
A big drawback is that TransE struggles with one-to-many, many-to-one and many-to-many relations. 
TransH, Translating Embeddings on Hyperplanes, was proposed by Wang et al. \cite{wang14transh} to address the limitations of TransE, particularly in handling complex relations. TransH models each relation as a hyperplane, allowing entities to have different embeddings for different relations.
TransH can overcome the limitations of TransE and can handle complex relationships by projecting entities onto relation-specific hyperplanes. This also allows more flexible representations of the entities.
Knowledge graphs are increasingly used in recommender systems to capture complex entity relationships \cite{reinanda20knowledge}. 

The approach we present in this paper aims at developing a personalized retrieval pipeline that leverages a combination of knowledge graph to generate user models and of neural retrieval models. This combination captures both the semantic and structural aspects of academic documents, providing a novel and robust framework for enhancing personalized academic retrieval.

\section{Personalized Academic Retrieval with Knowledge-graphs}
\label{sec:methodology}

In this Section, we first introduce the overall architecture of the proposed academic search system, which we will call Personalized Academic Retrieval with Knowledge-graphs (PARK). Then, we describe the proposed \textit{Retrieval} and  \textit{User Models} by detailing how they are designed and trained. Specifically, we focus on how we leverage a citation graph (CG) to create a Knowledge Graph (KG) that is then embedded in a learned latent space, where the KG entities are represented as high-dimensional vectors.

\begin{figure*}[t]
    \centering
    \includegraphics[width=\linewidth]{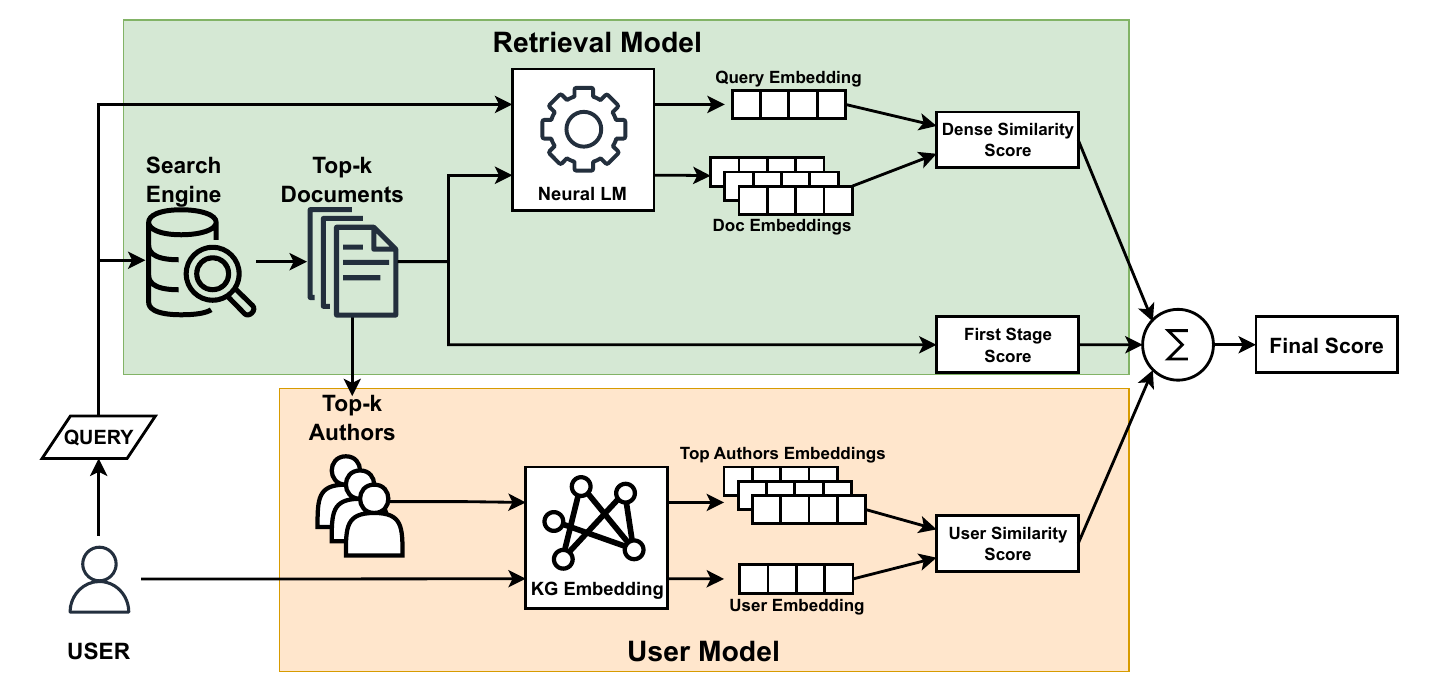}
    \caption{Overview of the PARK retrieval pipeline.}
    \label{fig:overview_model}
\end{figure*}

The overall architecture of PARK is illustrated in Figure \ref{fig:overview_model}. Our model has two main components: the first component, in the top part of the Figure, is a classical information retrieval IR pipeline that uses an efficient first retrieval stage based on BM25 and a following re-ranking stage exploiting our dense retrieval model. The second component, depicted in the lower part of the Figure, is the proposed user model based on knowledge graph embeddings, which computes user similarity scores, reflecting the alignment of research profiles and interests of the user issuing the query and the authors of the documents retrieved by the first stage retriever.

PARK integrates user models represented as embeddings derived from a knowledge graph’s latent representation, along with document similarity scores generated by a lexical model, in this paper we make use of BM25 \cite{robertson94okapi}. Additionally, it considers the semantic content of the same documents as represented in the dense neural retrieval model. The final documents presented to the user are ranked based on a weighted sum of three scores: BM25, dense similarity, and user similarity scores.

\subsection{The Retrieval Model}

We employ a two-stage retrieval pipeline to trade-off retrieval efficiency and effectiveness. In the first stage, we use lexical matching with BM25 to retrieve an initial set of documents for each query. 
In the second stage, re-ranking is performed using a state-of-the-art bi-encoder dense retrieval model.  
The retrieval pipeline outputs for each query-document pair two scores: the first stage score and the semantic similarity score computed by the neural re-ranker (See Figure \ref{fig:overview_model}).
The training of the neural retrieval model is performed using the procedure mentioned at the beginning of Section \ref{sec:related_work}. 
The neural retrieval model is optimized to minimize the distance between the query representation and the associated relevant document representations while increasing the distance between the query representation and the non relevant documents representations using the Triplet Margin Loss \cite{balntas16learning}, $\mathcal{L}(q,d^+,D^-)$, defined as: 
\begin{equation*}
    \sum_{d^- \in D^-} \max (||\mathbf{q}-\mathbf{d^+}||_p - ||\mathbf{q}-\mathbf{d^-}||_2 + \gamma, 0)
\end{equation*}
where $\mathbf{q}$ and $\mathbf{d}$ are the vector representation of the query and the document, respectively, $\gamma \in \mathbb{R}$ is the margin and $||\mathbf{q}-\mathbf{d^+}||_2$ is the norm of the difference between the vector representations of the query and of the document.

\subsection{The User Model: PARK-E and PARK-H}

We utilize the \textit{Closed World Assumption} (CWA) \cite{reiter81onclosed} when constructing our knowledge graph embeddings with both the TransE and the TransH algorithms. According to the CWA, any statement that is not known to be true is considered false. This assumption enhances the model’s ability to learn effectively from the available data. In this context, knowledge graph embeddings map entities and relationships into continuous vector spaces while preserving their structural and semantic characteristics. By applying the CWA, the embedding process allows the model to generate more accurate and robust representations.

\begin{figure*}[t]
    \centering
    \includegraphics[width=\linewidth]{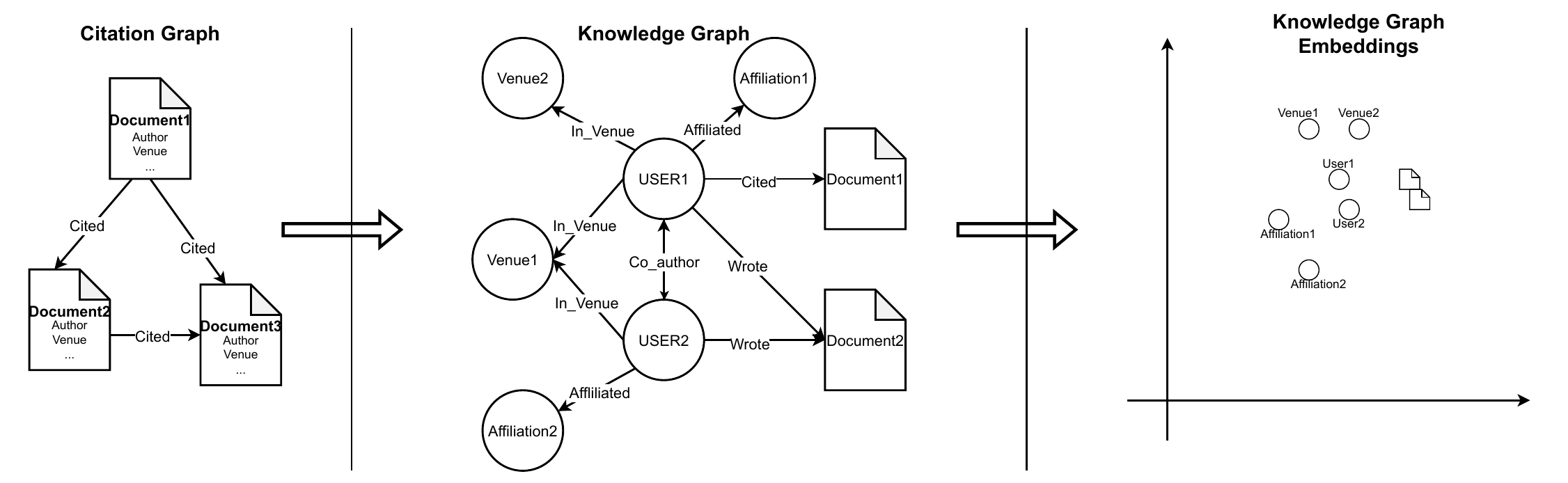}
    \caption{Overview of the process to obtain user embeddings from the citation Graph. The first step converts the Citation Graph to an Academic Knowledge Graph, and the second step embeds the Academic Knowledge Graph according to the proposed techniques.}
    \label{fig:academic_kg}
\end{figure*}

Our approach employs knowledge graph embeddings to represent various entities within the academic knowledge graph that we define, including users. The user embeddings are then used to score documents retrieved through the BM25 algorithm, based on the similarity between the user making the query and the authors of the retrieved documents. Below, we outline a two-step procedure, illustrated in Figure \ref{fig:academic_kg}, for learning the embeddings from the Citation Graph.

\paragraph{Academic Knowledge Graph Construction}
An academic citation graph is a directed graph that describes the citations within a collection of scientific documents. Each vertex (or node) in the graph represents a scientific document, and each edge in the graph is directed from one document to another, indicating that one document cites another. Additionally, each document node contains metadata, such as the author's name, affiliations, publication venue, and timestamp. This information can be used to transform the citation graph into a property graph.
A \textit{Property Graph} is a graph model where nodes are connected through edges that can have named properties. These graphs can be effectively used as knowledge graphs for embedding purposes, with each node and edge being considered in the embedding process. In our case, the nodes in the property graph represent papers, authors, institutions, or venues, and edges represent the relationships among these entities.
Academic citation graphs, to a reasonable extent, adhere to the CWA because, in these settings, the data is often curated and continuously updated, ensuring a high degree of completeness and accuracy.

To construct an academic knowledge graph from the academic citation graph, we define four types of nodes:
\begin{itemize}
    \item \textit{Users (or authors)}: These nodes represent authors of the academic documents.
    \item \textit{Documents}: These nodes represent scholarly works, including papers and articles, identified by their titles and abstracts.
    \item \textit{Venues}: These nodes denote the publication venues, such as conferences or journals, where scientific documents are published.
    \item \textit{Affiliations}: These nodes represent academic institutions, such as universities or research organizations, to which authors may belong.
\end{itemize}
These nodes can be connected by five types of relationships:
\begin{itemize}
    \item \textit{Wrote} (\( \text{user} \xrightarrow{\text{wrote}} \text{document} \)): Connects a user to all documents they have authored.
    \item \textit{Cited} (\( \text{user} \xrightarrow{\text{cited}} \text{document} \)): Connects a user to all documents cited by any of their authored works.
    \item \textit{In\_Venue}  (\( \text{user} \xrightarrow{\text{in\_venue}} \text{venue} \)): Connects the user to all venues where their documents were published.
    \item \textit{Affiliated}  (\( \text{user} \xrightarrow{\text{affiliated}} \text{affiliation} \)): Connects the user to their respective affiliation.
    \item \textit{Co\_author} (\( \text{user} \xleftrightarrow{\text{co\_author}} \text{user} \)): Connects a user to other users with whom they co-authored a document. This is the only symmetric relationship among the five.

\end{itemize}

\paragraph{{PARK-E and PARK-H}}
The user models, PARK-E and PARK-H, utilize the TransE and TransH algorithms respectively, to generate user embeddings from knowledge graphs. We selected TransE and TransH for their simplicity and effectiveness in capturing relationships between entities within a continuous vector space, making them well-suited for our user modeling needs. While other KG embedding techniques, such as TransR \cite{lin15learning} or RotatE \cite{sun19rotate}, provide more complex representations, TransE and TransH offer a balanced approach between computational efficiency and model performance, an essential consideration given the dimension of academic citation graphs. The complete pipeline, starting from the citation graph, is shown in Figure \ref{fig:academic_kg}.

The embedding dimension is aligned with the semantic space dimension of the neural language model used by the second stage re-ranker, allowing document semantics to be integrated into the knowledge graph embedding space. Specifically, we fix the embeddings of document nodes and use the embeddings generated by the neural language model to represent the documents. This approach preserves the semantics of the documents while infusing them into the knowledge graph.

The embeddings of all other nodes follow the procedure outlined by knowledge graph embedding algorithms, TransE \cite{bordes13transe} and TransH \cite{wang14transh}. The translational nature of TransE and TransH ensures that entities with similar relationships and contexts are positioned closer together in the embedding space, which is particularly beneficial for user modeling. Although TransE has limitations in handling complex relationships, such as one-to-many or many-to-many, these challenges are mitigated in this context, as clustering similar users together is a desirable property for effective user modeling.

\subsection{Integration of Retrieval and User Models}

The final ranking of the top-k retrieved documents by BM25 is based on three distinct scores, as illustrated in Figure \ref{fig:overview_model}. The first score is the dense similarity score, derived from the neural language model used for re-ranking. This score assesses the contextual and semantic relevance of documents to the query. The second score is obtained from the first stage retriever, which extracts the candidate results from the index using BM25. Finally, the user similarity score measures the similarity between the user embeddings of the query writer and the authors of the documents being scored, reflecting the alignment of their research profiles and interests.
The documents are ultimately re-ranked using a convex combination of the three normalized scores: BM25 scores, dense similarity scores, and user similarity scores. We describe this in detail in the next section.

\section{Experimental Evaluation}
\label{sec:experimental_settings}
In the following sections, we will first describe the datasets used in our experiments to evaluate the effectiveness of PARK. Next, we will introduce the competing methods relevant to the academic search task we are addressing. Finally, we will discuss the results of extensive experiments conducted to assess the performance of PARK compared to the competing methods on the specified datasets. The code is made publicly available.\footnote{\url{https://github.com/pkasela/PARK-Personalized_Academic_Retrieval_with_Knowledge-graphs}}

\subsection{Datasets and experimental settings}
For our experiments, we utilize four datasets specifically designed for comparing models in the context of personalized academic search, as introduced in \cite{bassani22amulti}. Each dataset focuses on a different domain: Computer Science, Physics, Political Science, and Psychology. Each domain contains approximately 5 million documents, resulting in a total of over 18 million documents and 1.9 million training queries.
Queries in this semi-synthetic dataset are generated by processing the titles of academic papers, removing stop words, and applying a Krovetz non-destructive stemmer to ensure they closely resemble real-world search queries. For both the training and validation sets, as in \cite{bassani22amulti}, relevant documents for each query are identified as those cited by the paper and those appearing in the top 100 results from a BM25 search, where the query is the exact paper’s title.

To create a realistic evaluation scenario and prevent data leakage, the dataset is split chronologically. The training sets consist of documents published before specific cutoff dates (2016 for Political Science, 2017 for Computer Science and Physics, and 2019 for Psychology), while the test sets include documents published after these dates. 

As the bi-encoder dense retrieval, we rely on \textit{MiniLM} \cite{wang20minilm}, a distilled version of BERT. MiniLM was selected for its computational efficiency and strong performance, making it particularly suitable for large-scale IR tasks. We train MiniLM for 10 epochs, with a learning rate of $5 \cdot 10^{-5}$ and a batch size of $256$.
The Knowledge Graph embedding model is trained for $100$ epoch with a batch size of $16384$ and a learning rate of $10^{-3}$. In both cases, we use AdamW \cite{loshchilov17fixing} as the model optimizer.

\subsection{Competing methods}
The competing baselines include traditional and neural retrieval models, user models, and graph-based approaches. Below, we describe each baseline:
\begin{itemize} 
\item \textbf{BM25} \cite{robertson94okapi}: The probabilistic information retrieval model used as the recall-oriented first-stage ranker in this study. This baseline uses the score returned by our first-stage ranker.
\item  \textbf{MiniLM} \cite{wang20minilm}: The precision-oriented neural re-ranker based on the pre-trained language model MiniLM and fine-tuned as detailed in Section \ref{fig:overview_model}. This baseline uses the score returned by our second-stage ranker.
\item  \textbf{Mean} \cite{bassani22amulti}: This user model is defined as the average representation of all documents related to a user.
\item  \textbf{Attention} \cite{jiang20end}: A query-aware user modeling technique that uses the Attention mechanism \cite{bahdanau14neural} to weigh the contributions of user-related document representations in generating the user model.
\item \textbf{Self Citation}: This baseline considers the self-citation behavior of authors, where documents cited by their own authors (or co-authors) receive a boost in their relevance score. 
\item \textbf{CrossEnc$_{\text{RA}}$} \cite{salemi24lamp}: A cross-encoder model based on MPNet \cite{song20mpnet}, which inputs the query, the candidate documents and the top-1 document most similar to the query from the user-related information.
\item \textbf{CTRL$_{\text{It}}$} \cite{mysore2024memory}: This model relies on memory-augmented cross-encoders. It incorporates user profiles as editable memory representations derived from historical interactions, to control and adjust personalization.
\item \textbf{PageRank} \cite{page99thepagerank}:  A graph-based algorithm originally designed to rank web pages, which can also be applied to citation graphs to rank academic papers based on citation links. We use the NetworkX implementation of the PageRank algorithm \cite{langville05asurvey} with $\alpha$ set to 0.85. 
\item  \textbf{POP} (Popularity) \cite{bassani22amulti}: This baseline ranks documents based on their popularity, measured by the number of citations they have received. This simple heuristic assumes that more frequently cited documents are more relevant or important.
\end{itemize}
The code for CTRL$_{\text{It}}$ and CrossEnc$_{\text{RA}}$ has not been publicly released, so we report the performance metrics as stated in their paper. For all the other baselines (except the first two), the final relevance score for a document is computed by linearly combining the normalized scores from BM25, the neural re-ranker (MiniLM in our case), and the specific baseline method using a convex sum \cite{mitra17learning}. This approach harnesses the strengths of traditional lexical matching (BM25), semantic understanding (MiniLM), and the unique properties of each baseline to enhance retrieval performance \cite{kasela23sepef, kasela24sepqa}. The final score is expressed as follows:
\begin{align*}
    S(q,d) = & \lambda_1 \cdot \text{BM25}(q,d) \\  
    & + \lambda_2 \cdot \text{MiniLM}(q,d) \\
    & + \lambda_3 \cdot \text{Baseline}(q,d)
\end{align*}
where $\lambda_1 + \lambda_2 + \lambda_3 = 1$ and $\text{Baseline}(q,d)$ denotes the score from the respective baseline method (Mean, Attention, Self Citation, PageRank, or POP). The $\lambda_i$ values are optimized based on performance metrics from the validation set. Additionally, all baseline models are optimized, and those that are trainable are trained end-to-end to ensure a fair comparison.

\subsection{Experimental results}
\label{sec:results}

\begin{table*}[htp]
    \centering
    \caption{Effectivess of PARK-E and PARK-H compared to the competing methods on the four datasets. The best-performing model is highlighted in boldface. Symbol * indicates a statistically significant difference over the second-best-performing model.}
    \label{tab:all_results_table}
    \resizebox{\textwidth}{!}{%
    \begin{tabular}{r|ccc|ccc|ccc|ccc}
         & \multicolumn{3}{c|}{Computer Science} & \multicolumn{3}{c|}{Political Science} & \multicolumn{3}{c|}{Psychology} & \multicolumn{3}{c}{Physics} \\
         \midrule
         Model &
         MAP@100 & MRR@10 & NDCG@10 & 
         MAP@100 & MRR@10 & NDCG@10 &
         MAP@100 & MRR@10 & NDCG@10 & 
         MAP@100 & MRR@10 & NDCG@10
         \\
         \midrule
         BM25 &
         0.123 & 0.489 & 0.225 & 
         0.133 & 0.502 & 0.241 &
         0.126 & 0.512 & 0.239 &
         0.128 & 0.537 & 0.269
         
         \\
         MiniLM &
         0.193 & 0.600 & 0.301 &
         0.186 & 0.580 & 0.297 &
         0.218 & 0.647 & 0.342 &
         0.183 & 0.624 & 0.335
         
         \\
         Mean &
         0.199 & 0.606 & 0.308 & 
         0.193 & 0.598 & 0.306 &
         0.220 & 0.652 & 0.347 &
         0.189 & 0.639 & 0.345 
         
         \\
         Attention &
         0.201 & 0.612 & 0.312 &
         0.199 & 0.612 & 0.314 &
         0.220 & 0.656 & 0.349 &
         0.190 & 0.648 & 0.348 
         
         \\
         Self Citation &
         0.213 & 0.624 & 0.325 &
         0.205 & 0.613 & 0.321 & 
         0.237 & 0.689 & 0.370 &
         0.204 & 0.671 & 0.365 
         
         \\
         CTRL$_{\text{It}}$ &
         - & 0.629 & 0.322 &
         - & 0.648 & 0.338 &
         - & 0.685 & 0.370 &
         - & 0.667 & 0.366
         
         \\
         CrossEnc$_{\text{RA}}$ &
         - & 0.635 & 0.324 &
         - & 0.651 & 0.338 &
         - & 0.700 & 0.380 &
         - & 0.673 & 0.369
         
         \\
         PageRank &
         0.213 & 0.644 & 0.331 &
         0.203 & 0.622 & 0.324 &
         0.230 & 0.670 & 0.360 &
         0.189 & 0.636 & 0.346
         
         \\
         POP &
         \textbf{0.238}* & \textbf{0.684}* & \textbf{0.370}* &
         0.214 & 0.649 & 0.345 &
         0.225 & 0.656 & 0.356 &
         0.206 & 0.670 & 0.370 
         
         \\
         \midrule
         PARK-E &
         0.228 & 0.651 & 0.344 &
         0.232 & 0.661 & 0.356 &
         \textbf{0.261}* & \textbf{0.716}* & \textbf{0.397}* &
         \textbf{0.225} & 0.695 & \textbf{0.391}
         
         \\
         PARK-H &
         0.230 & 0.655 & 0.346 &
         \textbf{0.233}* & \textbf{0.662}* & \textbf{0.357}* &
         0.255 & 0.712 & 0.392 &
         \textbf{0.225}* & \textbf{0.696}* & \textbf{0.391}*
         
    \end{tabular}%
    }

\end{table*}

Table \ref{tab:all_results_table} presents the results of our experiments, showcasing the performance of our proposed PARK-E and PARK-H models in comparison to various baseline methods discussed in the previous section. The evaluation covers four domains: Computer Science, Political Science, Psychology, and Physics. We used three evaluation metrics: Mean Average Precision@100 (MAP@100), Mean Reciprocal Rank@10 (MRR@10), and Normalized Discounted Cumulative Gain@10 (NDCG@10).
The findings indicate that the PARK-E and PARK-H models significantly outperform the baseline methods across nearly all domains. The only exception is in Computer Science, where the popularity-based (POP) baseline performs slightly better.

First, we note that all baseline models demonstrate improved retrieval effectiveness over the BM25 baseline. In the Computer Science domain, the PARK-H model shows a significant improvement of 8\% in MAP@100 compared to the self-citation user model; however, it falls short by 3\% when compared to the popularity-based (POP) baseline. This indicates that popularity, as measured by citation counts, plays a crucial role in determining the relevance of search results in this area.
In contrast, the PARK-H model consistently outperforms all other baselines in the domains of Political Science, Psychology, and Physics. Specifically, it improves MAP@100 over the second-best performing model by 9\%, 10\%, and 9\%, respectively. These improvements are statistically significant, as indicated by the asterisk in Table \ref{tab:all_results_table}, highlighting the robustness of our models.

Overall, these findings suggest that PARK-E and PARK-H are highly effective for personalized academic search, with the exception of the Computer Science domain, where the POP model exhibits slightly better performance. The success of our proposed models in most domains indicates their potential for broader application in academic information retrieval.

\section{Ablation Study on Node Types}
\label{sec:ablation}
To better understand the contribution of each node type in our model, we conducted an ablation study across the four benchmarks on PARK-H. The goal was to evaluate the effect of each node type (user, venue, and affiliation) and their associated relations on the model's retrieval performance.
We started by implementing the model with only user nodes and their three associated relations (authored, cited, and co-authored). Next, we added the venue node and venue relation to evaluate any additional impact. Finally, we incorporated the affiliation node, i.e., the complete setup used in PARK.

\begin{table*}[ht]
\centering
\caption{Ablation study results for each node type on four datasets.}
\label{tab:ablation_study}
\resizebox{\textwidth}{!}{%
\begin{tabular}{l|ccc|ccc|ccc|ccc}
 & \multicolumn{3}{c|}{\textbf{Computer Science}} & \multicolumn{3}{c|}{\textbf{Political Science}} & \multicolumn{3}{c|}{\textbf{Psychology}} & \multicolumn{3}{c}{\textbf{Physics}} \\ \hline
\textbf{Node Types} & \textbf{MAP@100} & \textbf{MRR@10} & \textbf{NDCG@10} & \textbf{MAP@100} & \textbf{MRR@10} & \textbf{NDCG@10} & \textbf{MAP@100} & \textbf{MRR@10} & \textbf{NDCG@10} & \textbf{MAP@100} & \textbf{MRR@10} & \textbf{NDCG@10} \\ \hline
Only User & 0.223 & 0.644 & 0.336 & 0.220 & 0.637 & 0.343 & 0.244 & 0.693 & 0.379 & 0.221 & 0.686 & 0.383 \\
\quad \reflectbox{\rotatebox[origin=c]{180}{$\Rsh$}} + Venue & 0.225 & 0.642 & 0.337 & 0.225 & 0.646 & 0.348 & 0.247 & 0.694 & 0.381 & 0.222 & 0.686 & 0.383\\
\qquad \reflectbox{\rotatebox[origin=c]{180}{$\Rsh$}} + Affiliation & 0.230 & 0.655 & 0.346 & 0.233 & 0.662 & 0.357 & 0.255 & 0.712 & 0.392 & 0.225 & 0.696 & 0.391
\end{tabular}%
}
\end{table*}

With \textbf{only user nodes} and the associated relations, we observe a notable improvement in retrieval performance across all datasets when compared to the baseline model without any other knowledge graph nodes. This improvement underscores the critical role of user-specific information in personalized academic search tasks. Adding the \textbf{venue node} and its associated relation yielded minimal performance gains in terms of NDCG@10. NDCG@10 increased by only 0.3\%, 1,5\%, 0.5\% in the Computer Science, Political Science, and Psychology benchmark. In Physics, NDCG@10 remained constant at 0.383. These limited improvements suggest that publication venue information contributes less significantly to user preference modeling, suggesting that venue context may not heavily influence individual user preferences within these academic domains.
The performance increases significantly across all datasets when the \textbf{affiliation node} is added, completing the full PARK model. NDCG@10 increases by 2.7\%, 2.6\%, 2.9\% and 2.1\% for Computer Science, Political Science, Psychology, and Physics respectively. These results highlight the significant positive impact of including the affiliation node, which likely captures valuable institutional and collaboration context, further enhancing the retrieval with user research interests.

The ablation study indicates that while user-specific information is the primary driver of performance gains, the affiliation node further enhances retrieval effectiveness by adding valuable institutional context. In contrast, the venue node has limited impact, suggesting it may be less essential for capturing user-specific preferences in academic search tasks.

\section{Conclusion}
\label{sec:conclusion}

The experimental results clearly demonstrate that our proposed user models, PARK-E and PARK-H, significantly outperform baseline methods across multiple academic domains, showing particularly strong performance in Political Science, Psychology, and Physics. Although the POP model slightly outperforms PARK models in the Computer Science domain, PARK remains the most effective user-centric model, underscoring the validity and robustness of our approach.
These findings highlight the substantial benefits of integrating knowledge graph embeddings into personalized information retrieval. By leveraging user embeddings derived from TransE and TransH algorithms, our models achieve more accurate document re-ranking by effectively capturing user similarities and preferences, thereby significantly enhancing retrieval performance.
However, these promising outcomes are accompanied by certain limitations. The effectiveness of knowledge graph embeddings is intrinsically linked to the quality and completeness of citation data. Incomplete or biased citation records can adversely impact embedding quality and, consequently, retrieval performance. Moreover, our model currently operates under a closed-world assumption, which, while suitable for academic search contexts, may restrict its applicability to other domains or tasks. This highlights the need for further exploration and adaptation to enhance generalizability.
Additionally, the decision to fix document nodes during training may constrain the model’s flexibility. Future research could explore the introduction of soft constraints in the loss function, allowing document nodes to adjust dynamically while still converging toward optimal document embeddings.
Other interesting research directions include exploring hybrid models that integrate popularity-based features with knowledge graph embeddings to improve retrieval effectiveness. Relaxing the fixed document node constraint could also offer a balance between flexibility and accuracy, potentially unlocking even greater performance gains.

In conclusion, the integration of knowledge graph embeddings into user modeling represents a significant advancement in personalized information retrieval. Our results demonstrate substantial improvements in retrieval performance across various academic domains, suggesting that these techniques have the potential to significantly advance the state of the art in academic search. With continued refinement and optimization, these methods are poised to make a lasting impact on the field of personalized information retrieval.

\section*{CRediT authorship contribution statement}
\textbf{Pranav Kasela:} Writing – original draft, Visualization, Validation, Software, Resources, Methodology, Formal analysis, Data curation, Conceptualization. \textbf{Gabriella Pasi:} Writing – review \& editing, Writing – original draft, Validation, Supervision, Methodology, Investigation. \textbf{Raffaele Perego:} Writing – review \& editing, Writing – original draft, Validation, Supervision, Methodology, Investigation.

\section*{Funding sources}
This work was supported by the European Union – Next Generation EU within the project NRPP M4C2, Investment 1.,3 DD. 341 - 15 march 2022 – FAIR – Future Artificial Intelligence Research – Spoke 4 - PE00000013 - D53C22002380006

This work was partially supported by the Spoke ``Human-centered AI'' of the M4C2 - Investimento 1.3, Partenariato Esteso PE00000013 - ``FAIR - Future Artificial Intelligence Research'', the ``Extreme Food Risk Analytics'' (EFRA) project, Grant no. 101093026, funded by European Union – NextGenerationEU.

\section*{Data availability}
The data used is already publicly available and the respective resource papers have been cited.



\bibliographystyle{elsarticle-num} 
\bibliography{biblio.bib}



\end{document}